\documentclass{article}
\usepackage[utf8]{inputenc}
\usepackage{graphics,graphicx,amssymb,amsmath,dcolumn,bm,url,amsbsy,amsfonts,amstext,pgf}
\usepackage[none]{hyphenat}
\usepackage{microtype}
\usepackage{setspace}
\usepackage{geometry}
\usepackage{pslatex} 
\usepackage{indentfirst}
\usepackage{array}
\usepackage{booktabs}
\usepackage{authblk}
\date{\today}

\begin{document}
\title{Shearless effective barriers to chaotic transport induced by even twin islands in nontwist systems}
\author[1]{M. Mugnaine \thanks{mmugnaine@gmail.com}}
\author[1]{I. L. Caldas}
\author[2,3]{J. D. Szezech Jr.}
\author[4]{R. L. Viana}
\author[5]{ P. J. Morrison}
\affil[1]{Institute of Physics, University of S\~ao Paulo, S\~ao Paulo, SP, 05508-090, Brazil}
\affil[2]{Graduate in Science Program—Physics, State University of Ponta Grossa, Ponta Grossa, PR, 84033-240, Brazil}
\affil[3]{Department of Mathematics and Statistics, State University of Ponta Grossa, Ponta Grossa, PR, 84033-240, Brazi}
\affil[4]{Department of Physics, Federal University of Paran\'a, Curitiba, PR, 81531-980, Brazil}
\affil[5]{Department of Physics and Institute for Fusion Studies,  The University of Texas at Austin, Austin, TX 78712, United States}
\maketitle
\begin{abstract}

For several decades now it has been known that systems with shearless invariant tori, nontwist Hamiltonian systems,  possess barriers to chaotic transport.  These barriers are resilient to breakage under perturbation and therefore regions where they occur are natural  places to look for barriers to transport.  Here we describe a novel kind of effective barrier that persists after the shearless torus is broken.  Because phenomena are generic, for convenience we study the Standard Nontwist Map (SNM), an area-preserving map that violates the twist condition locally in the phase space.  The novel barrier occurs in  nontwist systems  when twin even period islands are present, which happens for a broad range of parameter values in the SNM.   With a phase space composed of regular and irregular orbits, the movement of chaotic trajectories is hampered by the existence of both shearless curves, total barriers, and a network of partial barriers formed by the stable and unstable manifolds of the hyperbolic points.  Being a degenerate system, the SNM has twin islands and, consequently, twin hyperbolic points.  We show that the structures formed by the manifolds intrinsically depend  on period parity of the twin islands.  For this even scenario the novel structure, named a torus free barrier (TFB), occurs because the manifolds of different hyperbolic points form an intricate chain atop a dipole configuration and the transport of chaotic trajectories through the chain becomes a rare event. This structure impacts the emergence of transport, the escape basin for chaotic trajectories, the transport mechanism and the chaotic saddle.  The case of odd periodic orbits is different:    we find for this case the  emergence of transport immediately after the breakup of the last invariant curve, and this leads to a scenario of higher transport, with intricate escape basin boundary and a chaotic saddle with non-uniformly  distributed points.

\end{abstract}

 
\section{Introduction}
\label{sec:intro}

The analysis of a Hamiltonian system can be reduced to area preserving maps, symplectic maps that arise as canonical Poincaré maps for systems of differential equations \cite{moser,meiss1992}. Near-integrable Hamiltonian systems possess a phase space with  coexistence of regular and irregular (chaotic) orbits. The questions of where chaotic trajectories go and how long they take to move from one region of the phase space to another naturally emerge, and are the subject of a large literature on transport theory  (see e.g. \cite{mackay1984,meiss2015}).   Transport theory concerns  the collective movement of chaotic trajectories and includes techniques for estimating the transport rates, which   can be related to  quantities in experiments  \cite{rom1990}. Transport analysis has application in many research fields, such as plasma confinement \cite{mackay1984transport,horton1998}, fluid dynamics \cite{pjmD93b,del1993},  and celestial mechanics \cite{koon2000,jorba2020}.

Hamiltonian systems possess an  intermixture of periodic, quasi-periodic,  and chaotic orbits and the movement of the irregular components can be affected by  partial barriers and remnants of invariant tori, the Cantori \cite{mackay1984,meiss2015}. A Cantorus is an invariant set with irrational frequencies formed by an infinite number of gaps \cite{mackay1984transport}, which can  provide a robust partial barrier in the phase space. The existence of Cantori in the phase space has  only been proven for nondegenerated systems,  {i.e.}, systems where the frequency of the motion on a torus is related with the action of the torus by a monotonic function (see e.g.\ \cite{del1993,escande2018}).  Maps with this monotonic behavior are called twist maps, and the KAM theorem and Aubry-Mather theory can be applied since the twist condition, defined for maps as ${\partial x_{n+1}}/{\partial y_n} \ne 0$, is satisfied at  every point in the phase space \cite{del1996,morrison2000}. A map that violates such condition is called nontwist and its universal behavior can be described by the standard nontwist map (SNM) given in \cite{del1993}, the simplest mathematical model that violates the twist condition. 

Even though the existence of Cantori  has not been  proven for nontwist maps and conventional  KAM theorem cannot be applied at every point of the phase space, there are rigorous KAM results for nontwist maps \cite{delshams00}.  Moreover,  transport has been widely studied  in nontwist systems (e.g. \cite{mugnaine2018,mugnaine2020,viana2021,szezech2009,szezech2012,grime2023,caldas2012,horton2009,tigan2016}). In the SNM, depending on parameter values, the phase space can be  composed of a chaotic sea bounded by invariant curves with two chains of islands immersed in the chaotic regions. Due to the violation of the twist condition, the two chains of islands are twins, {i.e.}, have the same frequency but are not connected \cite{egydio1992,oda1995}. Among the invariant curves, there is the shearless curve formed by the points where the twist condition is violated. The shearless curve is also the extremum point of the nonmonotonic frequency function. In a transport portrayal, the shearless curve acts as a barrier, {i.e.}, divides the chaotic sea into two regions and thereby preventing  global transport. Thus, the transition to global chaos is related to the destruction of this barrier as was first investigated by Greene's method in \cite{del1996}.

After the shearless curve breaks,   chaotic orbits can be limited by the existence of partial barriers. As noted in \cite{szezech2009}, there can be  an effective barrier formed by the remnants of the shearless curve,  which trap the chaotic orbits. There is also the phenomenon of stickiness around the twin islands. The manifolds also have an important role in creating the partial barriers \cite{corso1998,mugnaine2018}, since the intra-intercrossing scenario of the manifolds has a effect on the transport intensity. Interestingly, the transport studies of these earlier works were performed for the standard nontwist map with island chains of odd period. Due to the map symmetries, the relative position of the islands change for different parity.  For the odd case, a  hyperbolic and an elliptic orbit are aligned, while  for the even case two elliptic points are aligned. Because there is no dynamical manifold based description for the scenario of partial barriers with islands of even period, we provide numerical evidence for characterizing the behavior.  Our characterization of these partial barriers yields an  explanation for zero and low transport for certain parameters of the SNM.

In our numerical survey of the SNM we analyze even and odd scenarios and compare the transport mechanisms  acting in the two cases. We show that for the even scenario, the transport of chaotic trajectories is obstructed by a network formed by the stable and unstable manifolds related to the chains of hyperbolic points. We called this network structure a torus free barrier (TFB), since there is no invariant curve, but the transport rarely happens due to the manifold configuration. These results can be generalized for more complex degenerate Hamiltonian systems with the same structures of twin island chains.

Our paper is organized as follows: in Sec.\  \ref{sec:parity} we present the standard nontwist map and discuss features of the phase space related to the parity of the islands. We also discuss the nonmonotonic behavior of the frequency function. Transmissivity through the phase space is discussed in Sec.\ \ref{sec:TransX}.  Here we investigate the escape of  chaotic trajectories from  the escape basin and measure their escape times. In Sec.\ \ref{sec:TFB}, we describe the TFB structure that is responsible for the null/low transport of the system. Then, in Sec.\ \ref{sec:TM} we further analyze the transport through the TFB, explaining the rarity of such transport of a chaotic trajectory.  Finally, in Sec.\  \ref{sec:concl} we conclude.

\section{Parity scenarios for the standard nontwist map}
\label{sec:parity}

The standard nontwist map is an area preserving map that was obtained from a Hamiltonian model that  describes transport and mixing properties of Rossby waves in zonal shear flows \cite{del1993}.  It also arises from a model for the motion of charged particles in drift waves in confined plasmas \cite{horton1998}. The map is given by
\begin{eqnarray}
\begin{aligned}
    y_{n+1}=&y_n-b\sin(2 \pi x_n),\\
    x_{n+1}=&x_n+a(1-y_{n+1}^2), \hspace{2em} \text{mod}~~ 1,
    \label{eqSNM}
\end{aligned}
\end{eqnarray}
where the canonically conjugate variables are $y\in \mathbb{R}$ and $x \in [0,1]$.  The parameters $a$ and $b$ are real  and independent of each other,  and the domain of interest is $a\in[0,1]$ and $b\in \mathbb{R}$.   The map violates the twist condition, $\partial x_{n+1}/\partial y_n\ne 0$, at points in the phase space which belong to the so-called \textit{shearless curve}.  This  map was first  thoroughly studied in \cite{del1996,del1997}, where the breakup and scaling of the shearless curve were analyzed.

For $b\neq0$  the SNM represents  a nonintegrable Hamiltonian system with a phase space with coexistence of chaotic and regular solutions. Quasiperiodic solutions, such as the shearless curve, can be represented by a curve that spreads across the whole domain of $x$. There are also quasiperiodic orbits that wind around  any stable periodic solution. These  orbits constitute the islands and, due to the violation of the twist condition and the symmetry of the map, they come in pairs adjacent to  the shearless curve \cite{del1997,petrisor2001}. 

The search for periodic solutions can be reduced to a one-dimensional problem by the use of symmetries of the system, as first explained and used in \cite{del1996}. In the SNM, the primary periodic stable orbits (elliptic points) are always on one of the symmetry lines, defined by
\begin{eqnarray}
\begin{aligned}
    s_0=&\{(x,y)|x=0\},\\
    s_1=&\left\{(x,y)|x= {a(1-y^2)}/{2}\right\}.
    \label{eqSL}
\end{aligned}
\end{eqnarray}
Thus, the search for stable periodic orbits is a root finding problem along these lines. The symmetry properties of the SNM are due to the symmetry transformation $T_S(x,y)=\left(x+\frac{1}{2},-y\right)$, with which the relation $M_{SNM} T_S = T_S M_{SNM}$, with $M_{SNM}$ defined by (\ref{eqSNM}), is valid. The symmetry transformation can be applied to the symmetry lines in (\ref{eqSL}), and  two other lines can be found:
\begin{eqnarray}
\begin{aligned}
    s_2=&\left\{(x,y)|x={1}/{2}\right\},\\
    s_3=&\left\{(x,y)|x={a(1-y^2)}/{2}+{1}/{2}\right\}.
    \label{eqSL2}
\end{aligned}
\end{eqnarray}

\begin{figure}[!h]
	\begin{center}
		\includegraphics[width=0.75\textwidth]{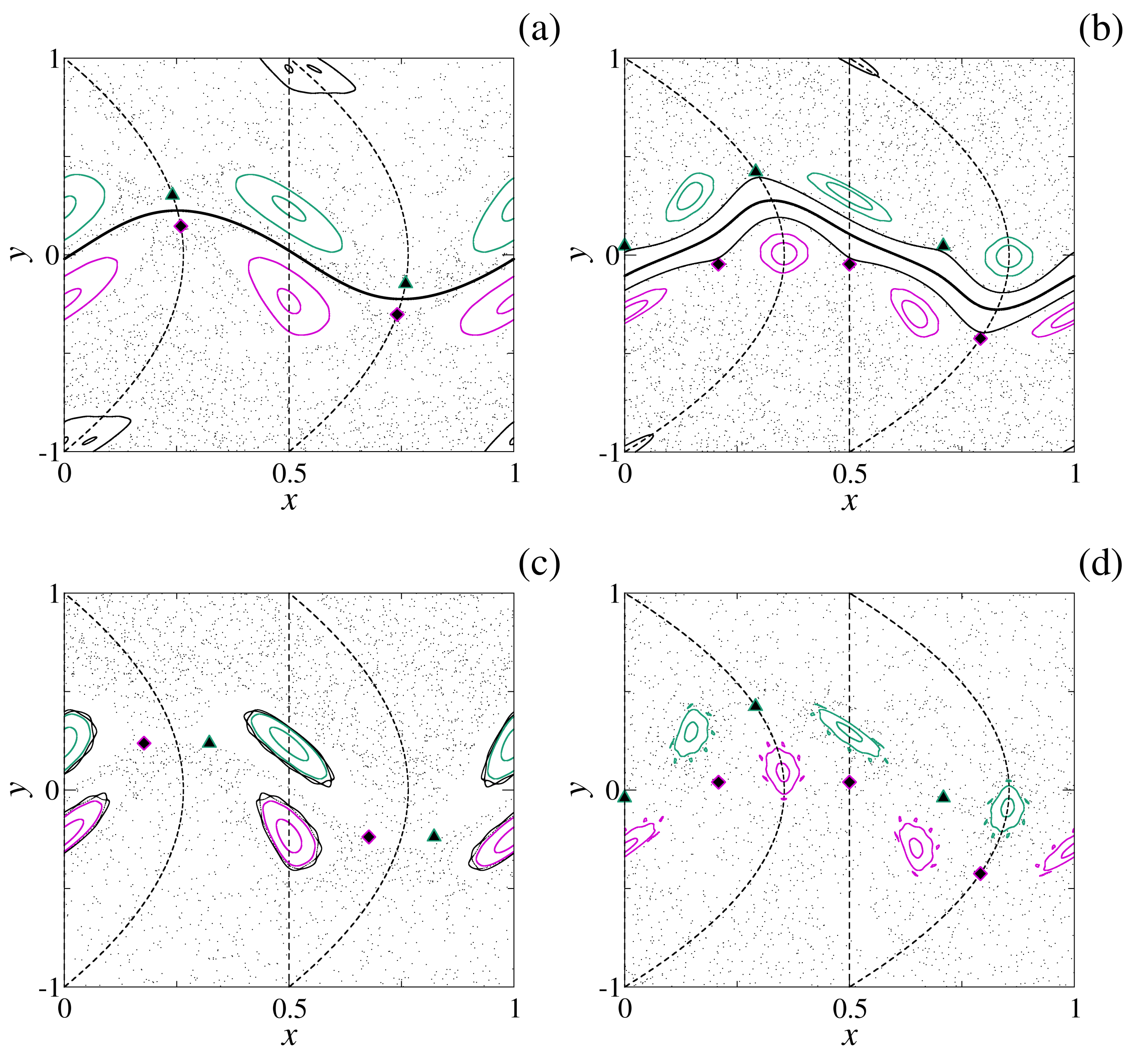}		
		\caption{Phase space plots  for the even and odd scenarios. In (a) and (c) we choose $a=0.53$ and for (b) and (d), $a=0.71$. The islands are represented by the colored closed curves, the hyperbolic points by the triangle and diamond symbols, and the shearless curve, when it exists, by the thick curve. The even scenario before the breakup of the shearless curve is shown in (a) for $b=0.45$, while the odd scenario with the curve is represented in (b), for $b=0.39$. In (c) and (d), we choose $b=0.53$ and $b=0.48$ in order to represent the even and odd scenario after the breakup of  the shearless curve, respectively.}
		\label{fig1}
	\end{center}
\end{figure}

As mentioned, the periodic orbits in the map come in pairs,  {i.e.}, there are two periodic solutions with the same frequency, one on each side of the shearless curve \cite{del1997}. According to the period parity of the periodic orbits, their relative positions are different if the period is even or odd. For a scenario where the periodic orbits exhibit even period, the fixed points of the same stability, of each chain of islands, belongs to the same symmetry line, while for odd period, the points with opposite stability are aligned along the symmetry line. In order to illustrate this, we compute  phase space plots for the SNM with islands of period 2 and 3, along with the four symmetry lines defined by (\ref{eqSL}) and (\ref{eqSL2}). The plots are shown in Figure \ref{fig1}.

In Figure \ref{fig1} we have phase spaces plots for the even and odd scenarios along with the symmetry lines $s_i$, for $i=0, 1, 2$ and 3 from equations (\ref{eqSL}) and (\ref{eqSL2}). The symmetry lines $s_0$ and $s_2$ are the vertical-dashed lines at $x=0$ and $x=0.5$ respectively, while the lines $s_1$ and $s_3$ correspond to the left and right dashed parabolas. For the even scenario,  we chose $a=0.53$ and $b=0.45$ for panel (a) and $b=0.53$ for panel (c). In the phase space plots  for the odd scenario, we have $a=0.71$ and $b=0.39$ for panel (b) and $b=0.48$ for panel (d). 

In Figure \ref{fig1} (a), the islands have period two, and they are indicated by the colored closed curves. We indicate the islands as upper and lower island chains with different colors: the upper (lower) chain  in green (magenta) is in the upper (lower) side of the shearless curve, represented by the thick black curve in the center. In panel (a), we observe elliptic points on the same symmetry lines $s_0$ and $s_2$ while the hyperbolic points, represented by the triangle and diamond symbols, are also in the same line, but in this case $s_1$ and $s_3$. The scenario for odd period is quite different. As seen in Figure \ref{fig1} (b), in the same symmetry line we have an elliptic and a hyperbolic point.

When the shearless curve no longer exists, we have the phase space plots of Figure \ref{fig1} (c) and (d), for the even and odd scenarios, respectively. In Figure \ref{fig1} (d), we observe the same behavior seen in Figure \ref{fig1} (b) when the curve was present: the elliptic and hyperbolic points are aligned along the symmetry line. This is not observed for the even scenario. As long as the elliptic points are still aligned in Figure \ref{fig1} (c), the hyperbolic points are out of the symmetry lines and, consequently, the two chains of hyperbolic points are no longer related by the symmetry transformation $T_S$. The hyperbolic points are out of the symmetry line due to the scattering that happens after the collision of the points at the shearless breakup. For a detailed discussion about the collision, see   \cite{del1996}.

\begin{figure}[!h]
	\begin{center}
		\includegraphics[width=0.8\textwidth]{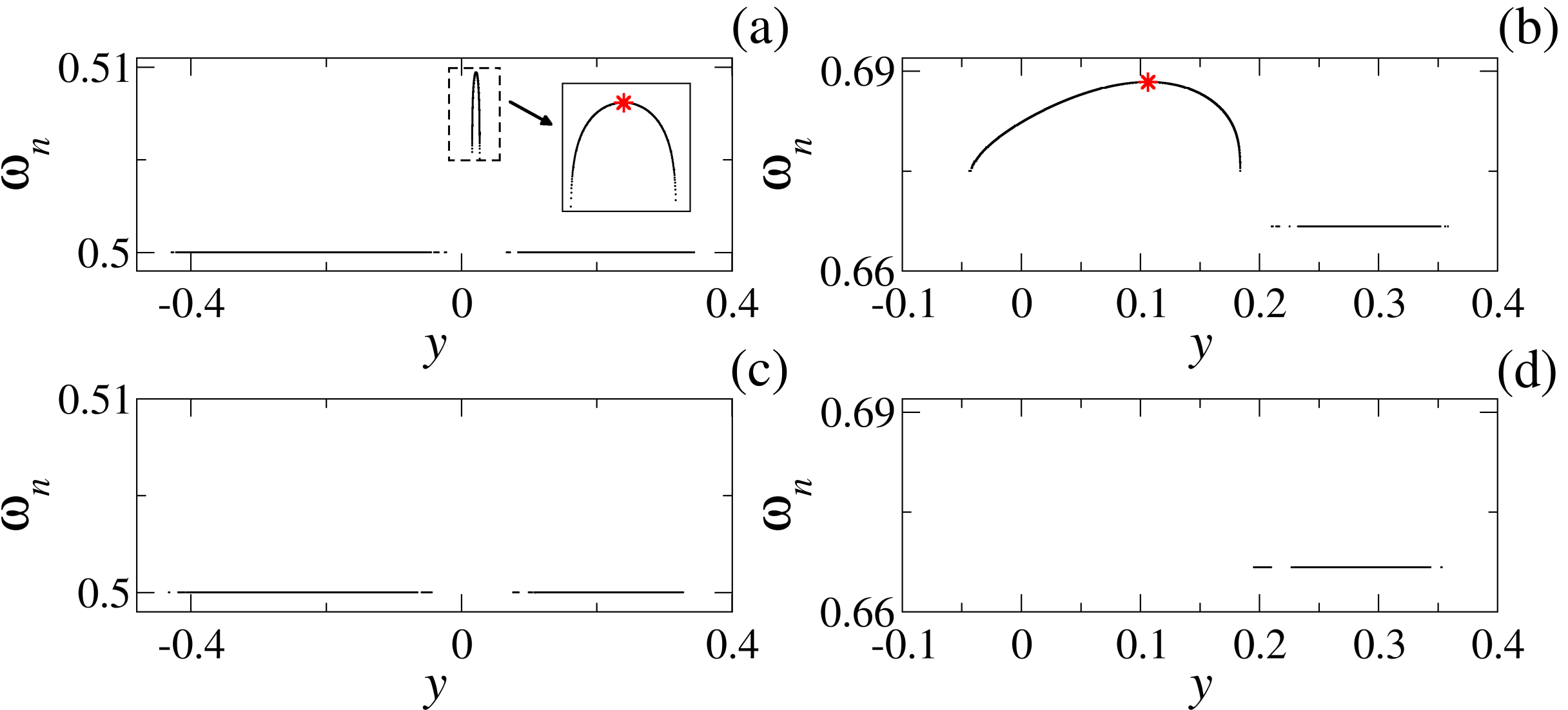}		
		\caption{Winding number profiles computed along the symmetry line $x=0.5$ for the phase space plots of  Figure \ref{fig1}. The parameters $a$ and $b$ for (a)-(d) are the same as those  for Figure \ref{fig1} (a)-(d), respectively. The red stars indicate the maximum value of $\omega_n$ in (a) and (b). The magnification in (a) is around the region framed by the dashed box.}
		\label{fig2}
	\end{center}
\end{figure}

The thick curves in Figure \ref{fig1} (a) and (b) are the shearless curve, the set of points where the twist condition is violated and the winding number assumes an extreme value. The winding number of an orbit, also called rotation number, is defined by the limit,
\begin{eqnarray}
    \omega_n=\lim_{n\to \infty} \dfrac{x_n}{n},
    \label{eqwn}
\end{eqnarray}
which converges to a rational (irrational) value for a periodic (quasiperiodic) orbit. The limit (\ref{eqwn}) does not converge if the orbit is chaotic. Computing the winding number profiles along the symmetry line $s_2: x=1/2$ for the phase space plots of Figure \ref{fig1}, we have the profiles shown in Figure \ref{fig2}.

For the winding number profiles of  Figure \ref{fig2}, we compute the limit (\ref{eqwn}) for $10^5$ initial conditions on  the symmetry line $x=1/2$ with a final iteration time $n=10^6$. Observing the phase space plots  of Figure \ref{fig1}, the majority of the solutions at the symmetry line are chaotic  and they do not have a defined $\omega_n$. For this reason, the domain of $y$ used for the winding number profiles are reduced to $y \in [-0.4,0.4]$ for Figure \ref{fig2} (a) and (c), the even scenario, and $y\in [-0.1,0.4]$ for Figure \ref{fig2} (b) and (d), the odd scenario.

The even scenario with shearless curve is shown in Figure \ref{fig2} (a), with  $a=0.53$ and $b=0.38$. The profile $\omega_n$ is composed of  two plateaus  at $\omega_n=0.5$ indicating the two islands of period 2. Between these two plateaus, we observe nonmonotonic behavior of the winding number,  highlighted by the magnification: a parabola with the extreme point indicating the position of the shearless curve. The points on the left and right of the maximum point are due to other curves that  occupy the whole domain of $x$ and are not shown in Figure \ref{fig1} (a). Similar behavior is observed in Figure \ref{fig2} (b), but in this case, we only observe one plateau at $\omega_n=1/3$, representing the upper island of period 3. This difference is due to the disposition of the islands in the symmetry lines as discussed before. Since the elliptic points are aligned along the symmetry line for the even case, we observe two plateaus, differently from the odd scenario where the hyperbolic point is aligned with the elliptic point and only one plateau can be seen.

Figure \ref{fig2} (c) and (d) represent the even and odd scenario  for $(a,b)=(0.53,0.53)$ and $(a,b)=(0.71,0.48)$, respectively. For these parameters, the shearless curve no longer exists and in both profiles, only the plateaus remain, indicating the persistence of the islands with the increasing of the perturbation amplitude $b$. Once again, we can observe two plateaus for the even case and just one for the odd scenario.

\section{Transport across the phase space}
\label{sec:TransX}

As observed in Figure \ref{fig1} (a) and (b), the shearless curve acts as a barrier to  transport, since it divides the chaotic sea into  upper and lower regions. Thus, the transition to global chaos and, consequently, global transport requires the destruction of the shearless barrier \cite{del1996}. For some set of parameters $(a,b$) where the shearless curve no longer exists, chaotic solutions can cross the phase space leading to  high transport. In order to quantify and describe the transport through the phase space, we use the transmissivity quantifier and the concepts of escape time and escape basins.

\subsection{Transmissivity}
\label{ssec:Trans}

\begin{figure}[!h]
	\begin{center}
		\includegraphics[width=0.9\textwidth]{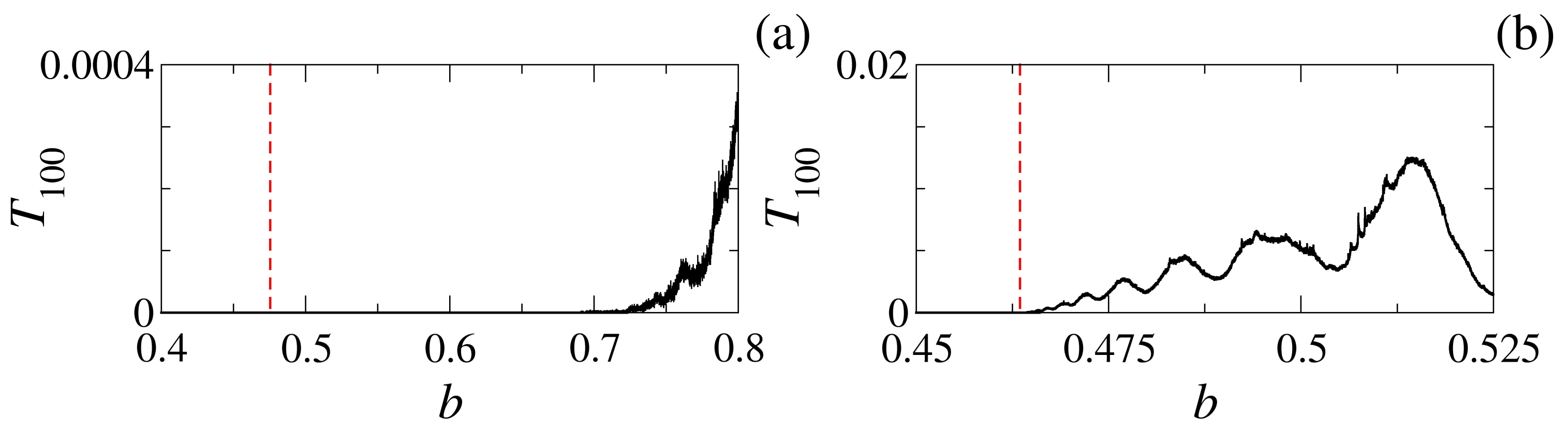}		
		\caption{Transmissivity profile for the SNM with (a) $a=0.53$ representing the even scenario and (b) $a=0.71$, representing the odd case. The red dashed line indicates the $b$  value for the  breakup of the shearless curve.}
		\label{fig3}
	\end{center}
\end{figure}

The transmissivity is a quantifier defined as the fraction of orbits that cross a region of  the phase space in a predefined time interval \cite{szezech2009}. Numerically, we randomly choose $N_T=10^6$ initial conditions on the line $y=-2.0$, iterate all of them for $100$ iterations and compute how many cross the phase space and reach the line $y=2.0$. The fraction is the transmissivity $T_{100}$. As performed before, we compute the quantifier for two scenarios: even and odd. For the even scenario we again choose $a=0.53$ and for the odd, $a=0.71$. The behavior of $T_{100}$ relative to the perturbation parameter $b$ is shown in Figure \ref{fig3}. 

From the transmissivity profiles in Figure \ref{fig3}, we can affirm that  transport across phase space with islands of even period differs from that with  islands of odd period. The $b$ values for the breakup of the shearless curve are indicated by the red dashed lines:  $b\approx 0.4754$ and $b\approx 0.4635$ for Figure \ref{fig3} (a) and (b), respectively. In the even scenario, represented by the profile in Figure \ref{fig3} (a), we observe that even without a barrier, the transport can be null for a large range of parameters for 100 iterations. The shearless curve breaks at the point $b\approx 0.4754$ but a non-null transport is noticed only for $b=0.70$. In contrast is the  result observed in Figure \ref{fig3} (b) where the islands in the phase space have odd period. In this case, the transport is non-null right after the shearless breakup. Therefore, for the even scenario, we have a transport barrier without an actual barrier, a TFB.

In order to investigate the difference between no/low transport for the even case and the  higher transport for the odd case, we analyze the phase space and the chaotic solutions for some values of $b$. We choose two values of $b$ for the even case, one for the scenario with a TFB and another for the higher transport observed in Figure \ref{fig3} (a), therefore, $b=0.53$ and $b=0.80$, respectively. For the odd case, we analyze the peak of transport for  $b=0.494$ of the transmissivity profile of Figure \ref{fig3} (b). We examine the three sets of parameters using the concepts of escape basin and escape time, stated in the next subsection. 

\subsection{Escape basins and escape time}
\label{ssec:EBET}

Consider now  how the chaotic solutions distributed over the phase space behave during the time evolution of the map. For this, we unite the escape time and the escape basin analysis, since we can set exits in the phase space of the SNM. The escape basin associated with an exit is the set of points which escape through this exit within a time interval, while the escape time is the number of iterations a solution takes to go through the exit. 

\begin{figure}[!h]
	\begin{center}
		\includegraphics[width=1.0\textwidth]{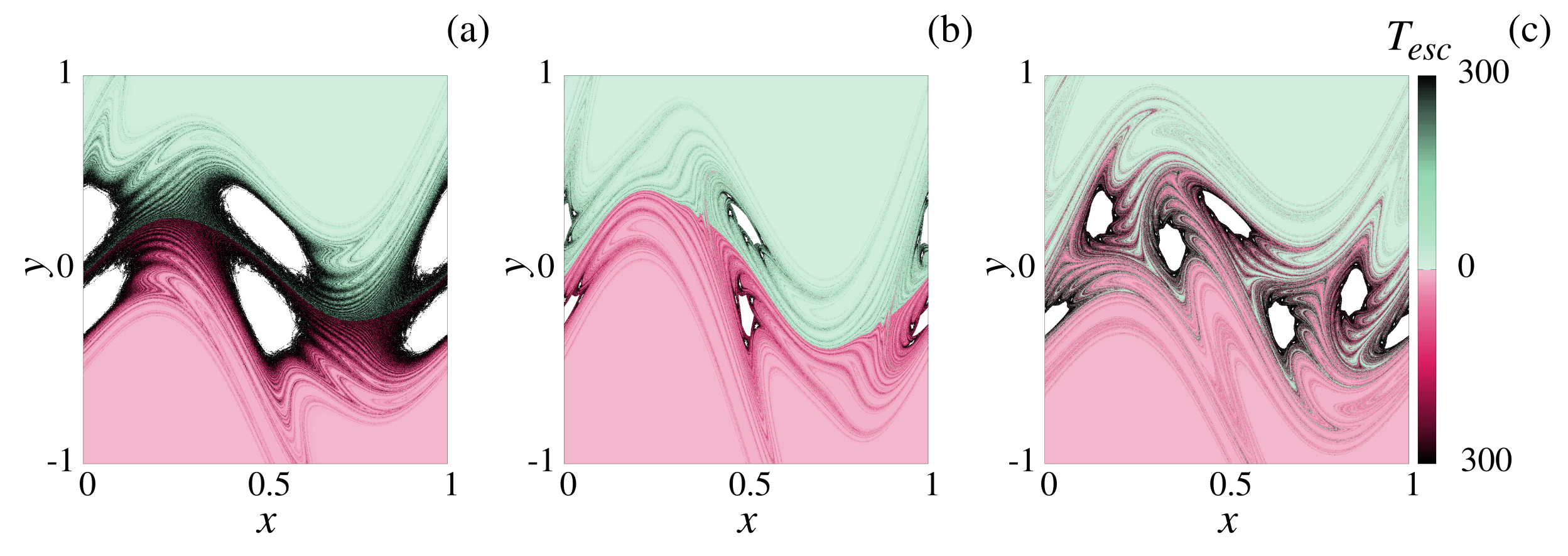}		
		\caption{Escape time and escape basin for the SNM for the even scenario with $a=0.53$, (a) $b=0.53$ and (b) $a=0.80$, and for the odd scenario, with (c) $(a,b)=(0.71,0.494)$. The pink (green) points indicate the initial condition of the solution which cross the exit $E^-: y=-1.0$ ($E^+: y=+1.0$) within $10^3$ iterations. The color shade, as shown in the color bar, indicate the escape time: light pink/green indicate a small escape time while the darker points indicate longer escape times.}
		\label{fig4}
	\end{center}
\end{figure}

As in previous  studies (see \cite{mugnaine2018,mugnaine2020,viana2021,mathias2019}), we define two exits  in the phase space: the lines $y=\pm 1.0$. Then we uniformly distribute $2. 10^3 \times 2. 10^3$ initial conditions in the domain $\mathcal{D}:\{(x,y)| x\in[0,1]~\text{and}~y\in [-1,1]\}$ and iterate all of them for $10^3$ iterations. During the time evolution, if the solution crosses one of the two exits $y=\pm1$ we record which exit and also the iteration number at which the crossing happens. We join these two pieces of  information in a color scale for the phase space. If the solution crosses the exit $E^+: y=+1$ ($E^-: y=-1$), we set its initial condition as a green (pink) point in the phase space. The shade of green and pink is related to the escape time: the darker it is, the longer the escape time. The escape basins and the escape times for $(a,b)=(0.53,0.53)$, $(a,b)=(0.53,0.80)$ and $(a,b)=(0.71,0.494$) are show in Figure \ref{fig4}.

Figure \ref{fig4} (a), with $(a,b)=(0.53,0.53)$ represents the even scenario with a TFB, while Figure \ref{fig4} (b) shows  the peak of transport for the even scenario of Figure \ref{fig3} (a), with $(a,b)=(0.53,0.80)$. For both cases of  Figure \ref{fig4} (a) and (b), we observe a smooth boundary between the green and the pink basins, {i.e.}, there is no significant mixing between the two colored regions. Thus, the trajectories which begin in the lower/upper region of the phase space tends to escape through the lower/upper exit and, consequently,   crossing   the phase space is a rare event. A different outcome is seen in the odd scenario of Figure \ref{fig4} (c), where the boundary between the basins is no longer smooth. In fact, the boundary is intricate and there is a mixing region of the pink and green points around the islands (the islands are indicated by the white regions). This mixing leads to a greater uncertainty, since close initial conditions can escape through different exits. From the result presented in Figure \ref{fig4} (c), we can also affirm that the crossing through the phase space is more common, comparing to the case in Figure \ref{fig4} (b), since there are green/pink points in the lower/upper region.

From Figure \ref{fig4}, we observe that greater escape times happen for trajectories with initial conditions near the islands. This is due to the stickiness phenomenon,  {i.e}, chaotic solutions get stuck around the quasi regular orbits for a certain time interval. Figure \ref{fig4} (a) presents the larger black region, indicating a stronger stickiness in this case. For Figure \ref{fig4} (b) and (c), we observe a more restrained stickiness region around the islands. The majority of the escape basins present light shades of pink and green, indicating a short escape time. The darker shades are closer to the islands and, apparently, distance themselves from the islands in patterns that resemble manifolds. Therefore, it seems the manifolds  reflect the behavior of the chaotic solutions and their transport through the phase space. 

 \section{Torus free barriers}
 \label{sec:TFB}

Transport analyses aim to characterize the motion of chaotic trajectories through phase space (e.g.\ \cite{meiss2015}). Typically nonintegrable Hamiltonian systems that are perturbations of integrable ones have  a coexistence of chaos and regularity, with islands and invariant curves that can affect, limit,  and/or preclude transport. The Poincaré-Birkhoff theorem states that for each elliptic fixed point at  the center of the islands there corresponds unstable hyperbolic fixed points in the chaotic sea. As shown in  \cite{mugnaine2018,szezech2009,szezech2012,corso1998}, the stable and unstable manifolds related to the hyperbolic points  impact  the transport of chaotic trajectories in the SNM. Following this idea in the present context, we give here a  manifold analysis for the phase space plots of Figure \ref{fig4},   in order to explain the torus free barrier scenario and the peaks of transmissivity through the phase space.

\begin{figure}[!h]
	\begin{center}
		\includegraphics[width=1.0\textwidth]{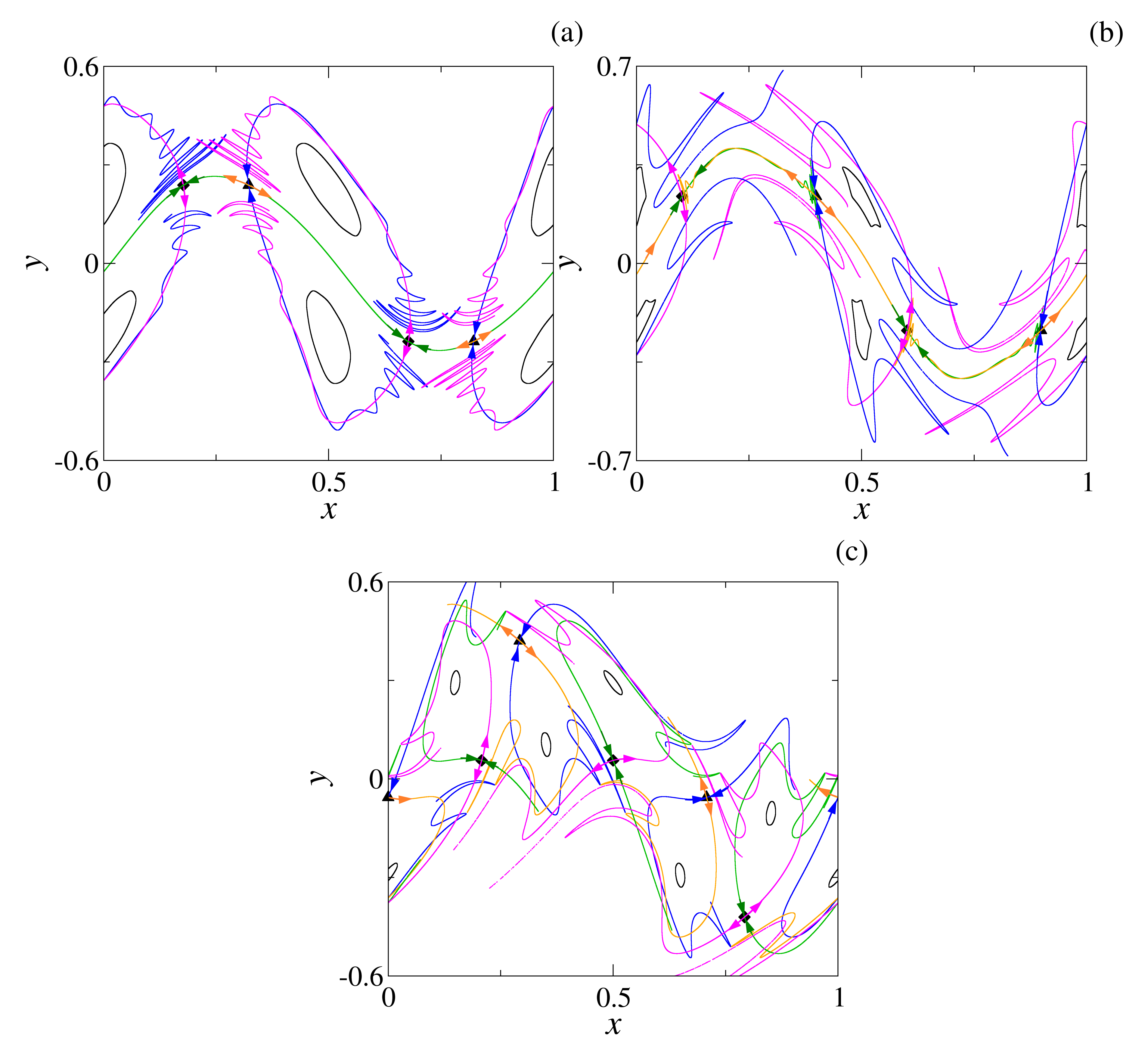}		
		\caption{Manifold structure for the phase space plots of Figure \ref{fig4}. The upper (lower) hyperbolic points $\mathbf{O^U}$ ($\mathbf{O^L}$) are represented by the triangle (diamond) symbol. The manifolds $W_s^U$, $W_s^L$, $W_u^U$ and $W_u^L$ are indicated by the  blue, green, orange and pink curves. Close to the hyperbolic points, we draw arrows indicating the stability direction: inwards and outwards, indicating  the stable and unstable directions of the hyperbolic point.}
		\label{fig5}
	\end{center}
\end{figure}

Following the notation of  \cite{mugnaine2018}, let $\mathbf{O^U} \left(\mathbf{O^L}\right)$ denote the hyperbolic points of the upper (lower) island chain and denote  its unstable and stable manifolds by  $W_u^U \left(W_u^L\right)$ and $W_s^U \left(W_s^L\right)$, respectively. Because  we are investigating  a scenario with  twin islands in nontwist systems, the crossing between stable and unstable manifolds can form an intracrossing scenario, where the crossings occur between manifolds of the same chain of hyperbolic points,   and  the intercrossing scenario, where the crossings are between the manifolds of different chains of hyperbolic points \cite{mugnaine2018,szezech2009}. For the phase space plots of Figure \ref{fig4}, the stable and unstable manifolds of each chain of hyperbolic points are shown in Figure \ref{fig5}.

Figure \ref{fig5} depicts  the stable and unstable manifolds related to  hyperbolic points of the SNM for the same parameters as those of Figure \ref{fig4}. In Figure \ref{fig5} (a)  the manifolds are shown  for the case of the TFB, where we have no transport without a real barrier. The blue and pink curves, representing   $W_s^U$ and $W_u^L$, respectively, surround the two islands in a dipole-like shape and the crossing structure between these two manifolds resembles  the usual turnstile/homoclinic structure. The intriguing and newly observed structure is the one formed by the green and orange curves, the stable and unstable manifolds of the lower and upper chain, $W_u^U$ and $W_s^L$, respectively.  These two manifolds are apparently superimposed and form what appears to be a  smooth curve across the interior of and between the dipoles. This smooth curve correspond to the smooth boundary between the two escape basins of Figure \ref{fig4} (a). From our numerical simulations, we affirm that there are no manifolds that connect  different hyperbolic points in the region inside and outside of the dipole. Even though it seems the manifolds occupy the same region;  in the region outside the dipole, the green curve oscillates and moves  away from the triangle hyperbolic point $\mathbf{O^U}$. The same happens with the orange curve: it gets closer to the diamond point $\mathbf{O^L}$, but then the manifold follows the pink manifolds and moves away from the point. In the curve inside the dipole, there are small oscillations and, consequently, crossings between the green and orange curves. (The crossings are shown in the magnification given in  Figure \ref{fig6}).

At the peak of transmissivity of Figure \ref{fig3} (a) we have $(a,b)=(0.53,0.8)$;  the corresponding  manifolds are presented in Figure \ref{fig5} (b). We observe a structure similar to the one presented in Figure \ref{fig5} (a), but the crossing between the orange and green curves are more evident. These crossings form lobes and enable  chaotic orbits to cross the phase space at these regions. Thus, the TFB structure is modified and the transport through the phase space is higher.

For the odd scenario, presented in Figure \ref{fig5} (c), we have the usual turnstile structure. This is a well-known case already investigated and described in the literature \cite{meiss2015,mugnaine2018}. The peak of transmissivity for this phase space structure  is due to the intercrossing between the manifolds of different chains of hyperbolic points. For a more complete description of how the intercrossing is responsible for an increase in the transport see   \cite{mugnaine2018}.

\section{Transport mechanism}
\label{sec:TM}

As noted above, the  manner in which  trajectories traverse phase space for the odd scenario case has been well-studied, including its understanding in terms of conventional  turnstiles. However,  to our knowledge,  the manner in which  trajectories cross the dipole-like  configuration for the even scenario case has not been previously investigated.  For the even case when TFBs exist,  because they are not perfect barriers, transport can occur;  however,  our numerical simulations show that this is rare  and the number of iterations needed is large. For example, for the phase space plot of Figure \ref{fig5} (a), fewer  than $0.01\%$ of the trajectories cross the phase space in $10^5$ iterations. Thus, the TFB is a  partial but quite strong transport barrier.

\begin{figure}[!h]
	\begin{center}
		\includegraphics[width=1.0\textwidth]{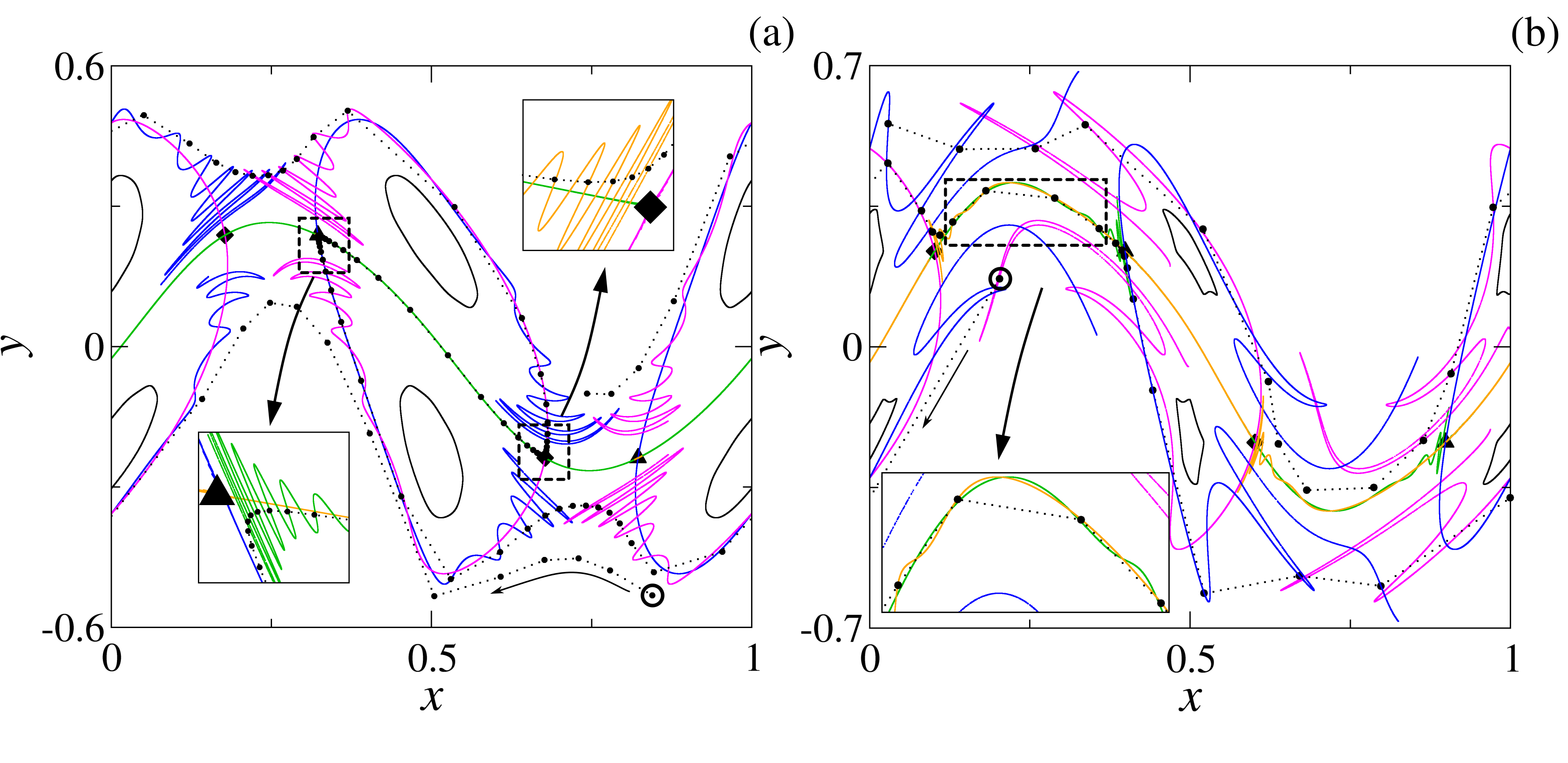}		
		\caption{Transport of a chaotic trajectory through the dipole-like configuration. The initial condition of the orbit is highlighted with a circumference and the direction of the orbit is indicated by  arrows. In (a), we have $b=0.53$ representing the scenario where the transport is a very rare event while in (b), $b=0.8$, we have the peak of transmissivity of the profile in Figure \ref{fig3} (a). The manifold  color scheme is the same as  of Figure \ref{fig5}. The insets in each panel are  magnifications of the corresponding regions inside the dashed squares. }
		\label{fig6}
	\end{center}
\end{figure}

In order to understand the transport mechanism, we search for trajectories that cross the phase space and analyze their behavior inside the dipole-like structure.  In particular, we analyze the transport through the structures shown in Figure \ref{fig5} (a) and (b). Our results are shown in Figure \ref{fig6},  where  we trace trajectory paths through the twin island chain region.  We consider representative initial conditions,  highlighted by circles,  and consider  the same parameters $a$ and $b$ of Figure \ref{fig5} (a) and (b), respectively. In Figure \ref{fig6} (a) an escape path is seen by  following the arrows indicating the direction of the trajectory.  We observe that the trajectory crosses  the TFB inside the dipole-like  configuration.  In contrast, in  Figure \ref{fig6} (b) we see the crossing happens in the region between the two dipoles. The magnifications of the insets show how the crossing happens: the trajectory goes from the lower part of the phase space to a region inside a lobe formed by the green and orange manifold, and then stays inside the lobes upon  each iteration. Then, due to the crossings between these manifolds, the trajectory goes to the upper region of the phase space. These scenaria occur for both phase space plots shown in Figure \ref{fig6}.

As discussed in \cite{mackay1984}, there is a relationship between the area of the lobes formed by the stable and unstable manifolds and the flux of trajectories through phase space. Therefore, we can explain the low/high transport in the phase spaces of Figure \ref{fig6} (a) and (b), respectively, by the area of the lobes formed by the manifolds. In Figure \ref{fig6} (a), the lobes formed by the green and orange manifolds are exceptionally small and can only be seen in the magnifications. Furthermore, they are restricted to the region inside the dipole region. The combination of small size with the restriction of being only inside the dipoles is  the reason the crossing of chaotic trajectories is a rare event in this configuration.   For the  different escape  path of  Figure \ref{fig6} (b), where the transport is higher, we can observe the lobes formed by the crossings without any magnification,  {i.e.}, they are significantly bigger in size. Moreover, there are significantly  more crossings in the regions between the dipoles, as highlighted by the magnification. In summary, the lobes between the dipoles boost the transport of chaotic trajectories in the phase space, due to the size and also the  increase in  the number of crossings of the manifolds.

As is well known , the stable and unstable manifolds are non-attracting invariant sets that play a fundamental role in the dynamics of the chaotic systems \cite{hsu1988}. The results of Figures \ref{fig5} and  \ref{fig6} support  this idea. Hence, we study a non-attracting invariant set as a  \textit{chaotic saddle}. Let $\mathcal{R}$ be a limited region in the phase space, in our case $\mathcal{R}$ is the domain $\mathcal{D}:=\{(x,y)| x\in[0,1]~\text{and}~y\in [-1,1]\}$. There is a set of points which remain in $\mathcal{R}$ for all time under forward iteration of the  ${SNM}$, and another set of points remain in the region for all time under the backward iteration of the map. These sets are the stable and unstable sets, respectively,  and their union is the chaotic saddle \cite{nusse1989}. We compute an approximation of the chaotic saddle by the ``sprinkler method" proposed in   \cite{hsu1988}. The method consists  of placing points on  a fine grid over the region $\mathcal{R}$ and iterating  all the points until a final iteration $n_F$.  The initial conditions of the trajectories that stay in the region during all the iterations  approximate the stable set, their final positions approximate the unstable manifold,  and the iteration ${n_F}/{2}$ approximates the chaotic saddle.

\begin{figure}[!h]
	\begin{center}
		\includegraphics[width=1.0\textwidth]{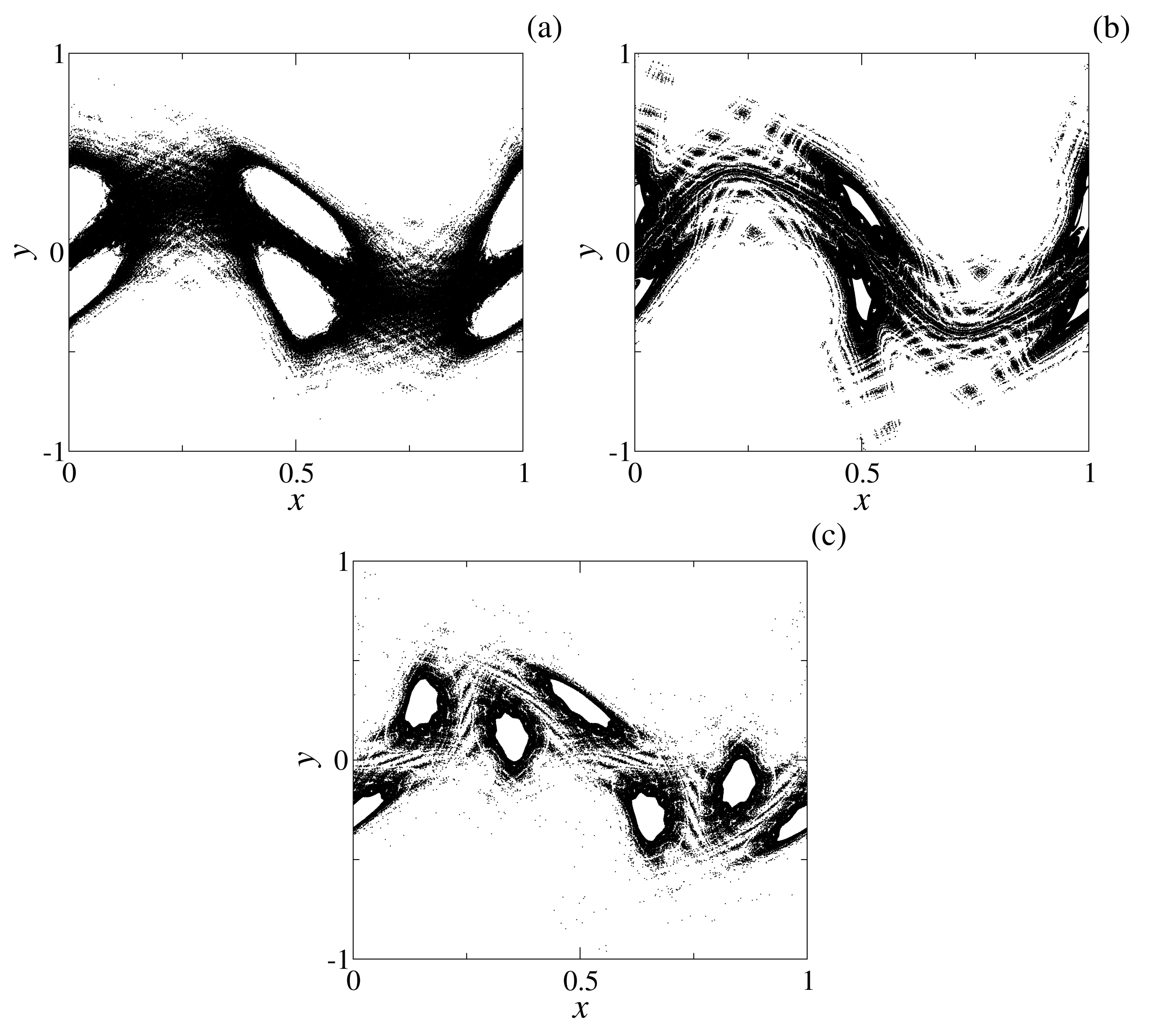}		
		\caption{Approximation of the chaotic saddle correspondent to the phase spaces of Figure \ref{fig5}. The computation is made in a grid of $2000 \times 2000$ initial conditions $n_F=100$. The black points represent the iteration $n=50$ for orbits that do not escape the region $\mathcal{D}:\{(x,y)| x\in[0,1]~\text{and}~y\in [-1,1]\}$ in 100 iterations.}
		\label{fig7}
	\end{center}
\end{figure}

To calculate the chaotic saddle for the phase space plots of Figure \ref{fig5}, we use a grid of $2000 \times 2000$ initial conditions in $\mathcal{D}$, discard   points that belong to the islands,  and consider the final iteration $n_F=100$. The chaotic saddle is show in Figure \ref{fig7}. From  Figure \ref{fig7} and the value of the transmissivity for each of the cases  of Figure \ref{fig5}, we can confirm that there is a relationship  between the spatial distribution of the invariant set and the transport through the phase space. The saddle of Figure \ref{fig7} (a) is related to the case with the TFB: here the chaotic saddle is dense and distributed in the whole domain of $x$. Visually, the density appears uniform on the whole saddle. A different scenario is observed in Figure \ref{fig7} (b). In this case, we observe that the chaotic saddle is  denser around the island of period 2 and the saddle has ``holes" and ``channels",  {i.e.}, paths for the chaotic orbits to cross the phase space with no restriction. Lastly, the chaotic saddle for the odd scenario, Figure \ref{fig7} (c), is dense around the islands but we can also observe the channels inside the saddle. With these results, it seems that a uniformly distributed and continuum saddle is related to a scenario with no transport through the phase space, while the saddles that are not uniformly distributed and are subdivided can lead to a scenario with non-null transport.

\section{Conclusions}
\label{sec:concl}

The transport of chaotic trajectories across the phase space is prevented in nontwist Hamiltonian systems by the existence of the shearless curve, an invariant set that is  robust to strong perturbations.  After the shearless curve breakup, the transport is affected by the structures formed by the stable and unstable manifolds related to the twin chains of hyperbolic points. We analyzed the transport in nontwist systems by using the paradigmatic standard nontwist map, and we observed the parity of the twin islands plays a crucial role in the collective motion of chaotic trajectories. For even period islands the transport can be null,  even though there is not a  complete barrier,  while for the odd scenario we observe the emergence of transport immediately after the shearless curve breakup.

Besides the transmissivity, the parity of the islands also affects the escape basin and the relative escape times. For the odd case, we observe an intricate boundary between the escape basins and   larger escape times are related to the stickiness and the hierarchy of the islands around islands structure. The boundary between the escape basin is smooth for the even case, and the escape times are related to the manifold structure that emerges in the even scenario case.

We justified the difference between the transport for the even and odd cases by examining the structures formed by the manifolds related to the chains of hyperbolic points. For the odd case, we  saw the  well-known structure of manifolds that can act like turnstiles for the chaotic trajectories. The main novelty of the present paper is the structure formed by the manifolds in the even scenario. In this case, we observed a dipole-like structure formed by the manifolds of different chains of hyperbolic points, characterizing an intercrossing scenario when  the period of the islands is even.

For the even scenario, the intercrossing always occurs, but this does not mean high transport, as in the odd case. The lobes formed in the intercrossing are small, decreasing the possibility of transport. The low/high transport can also be described by the chaotic saddle. The density and the continuity of the chaotic saddle play   important roles in the transport: dense and continuous saddles decrease the transport, while subdivided saddles facilitate the transport through phase space.

\section{Acknowledgments} 
This research received the support of the National Council for Scientific and Technological Development (CNPq - Grant No. 309670/2023-3, 403120/2021-7 and 301019/2019-3) and Fundação de Amparo à Pesquisa do Estado de São Paulo (FAPESP) under Grant No. 2018/03211-6, 2022/12736-0, and 2023/10521-0. PJM was supported by U.S. Dept. of Energy Contract $\#$DE-FG05-80ET-53088.

\end{document}